\documentclass[preprint2]{aastex}

\shorttitle{Long Term Quasar Variability. II}
\shortauthors{De Vries et al.}

\newcommand{\rsdss}{$r_{\mbox{\tiny SDSS}}$}
\newcommand{\gsdss}{$g_{\mbox{\tiny SDSS}}$}

\begin{document}
\title{Structure Function Analysis of Long-Term Quasar Variability}

\author{W. H. de Vries, R. H. Becker}
\affil{University of California, One Shields Ave, Davis, CA 95616}
\affil{Lawrence Livermore National Laboratory, L-413, Livermore, CA 94550}
\email{devries1@llnl.gov}

\and 

\author{R. L. White, C. Loomis}
\affil{Space Telescope Science Institute, 3700 San Martin Drive,
Baltimore, MD 21218}

\begin{abstract}
  
  In our second paper on long-term quasar variability, we employ a much larger
  database of quasars than in de Vries, Becker \& White. This expanded sample,
  containing 35\,165 quasars from the Sloan Digital Sky Survey Data Release~2,
  and 6\,413 additional quasars in the same area of the sky taken from the 2dF
  QSO Redshift Survey, allows us to significantly improve on our earlier
  conclusions. As before, all the historic quasar photometry has been calibrated
  onto the SDSS scale by using large numbers of calibration stars around each
  quasar position.  We find the following: (1) the outbursts have an asymmetric
  light-curve profile, with a fast-rise, slow-decline shape; this argues against
  a scenario in which micro-lensing events along the line-of-sight to the
  quasars are dominating the long-term variations in quasars; (2) there is no
  turnover in the Structure Function of the quasars up to time-scales of
  $\sim$40 years, and the increase in variability with increasing time-lags is
  monotonic and constant; and consequently, (3) there is not a single preferred
  characteristic outburst time-scale for the quasars, but most likely a
  continuum of outburst time-scales, (4) the magnitude of the quasar variability
  is a function of wavelength: variability increases toward the blue part of the
  spectrum, (5) high-luminosity quasars vary less than low-luminosity quasars,
  consistent with a scenario in which variations have limited absolute
  magnitude.  Based on this, we conclude that quasar variability is intrinsic to
  the Active Galactic Nucleus, is caused by chromatic outbursts / flares with a
  limited luminosity range and varying time-scales, and which have an overall
  asymmetric light-curve shape. Currently the model that has the most promise of
  fitting the observations is based on accretion disk instabilities.
  
\end{abstract}

\keywords{galaxies: active --- galaxies: statistics --- quasars:
general}

\section{Introduction}

The cause of the long-term variability in quasars is still a matter of debate.
Unlike the short time-scale variations (on the order of days), which are
adequately described in terms of relativistic beaming effects \citep[e.g.,][]
{bregman90,fan00,vagnetti03}, the variations at much longer time-scales (years
to decades) are less understood. Current scenarios under consideration are
ranging from source intrinsic variations due to Active Galactic Nucleus (AGN)
accretion disk instabilities \citep[e.g.,][]{shakura76,rees84,siemiginowska97,
kawaguchi98,starling04}, and possible bursts of supernovae events close to the
nucleus \citep[e.g.,][]{terlevich92,cidfernandes96}, to source extrinsic
variations due to micro-lensing events along the line-of-sight to the quasar
\citep[e.g.,][]{hawkins93, hawkins02,alexander95,yonehara99,zackrisson03}. See
also the review article by \citet{ulrich97}.

Determining which of the various proposed mechanisms actually dominates quasar
variability is best done by studying it toward the longest possible
time-baselines. Depending on the mechanism, each has markedly different
variability ``power'' at the longer time-scales \citep[e.g.,][]{hawkins02}. This
means that if one would have a quasar monitoring sample that is both large
enough, and covers a large enough time-baseline, one could address these issues
adequately.  Unfortunately, given the nature of monitoring programs, this is not
something that can be started overnight. The longest quasar light-curve
monitoring programs are on the order of 20 years \citep[e.g.,][]{hawkins96}, and
will take a long time before they are expanded significantly in time-baseline.

The way around this is by using historic photographic plate material, in
combination with a recent survey. Like in our previous paper \citep[hereafter
Paper~I]{devries03}, we chose to use the Sloan Digital Sky Survey (SDSS), Data
Release 2 (DR2), in combination with the historic Second Generation Guide Star
Catalog\footnote{The Guide Star Catalogue-II is a joint project of the Space
Telescope Science Institute and the Osservatorio Astronomico di Torino.}
\citep[GSC2,][]{mclean98} and the Palomar Optical Sky Survey
\citep[POSS,][]{reid91}.  This allows for photometric information on the quasars
spanning up to 50 years. The downside is that, unlike the monitoring programs,
we have typically a very sparse light-curve sampling {\it per quasar}. However,
since we will have a very large number of them, the sampling across the complete
database will be very good. This obviously only works if the variability of the
quasars is due to a mechanism common to all quasars. We proved the validity of
this concept in Paper~I, and recently a similar approach has been taken by
\citet{sesar04}.

The paper is outlined as follows: In \S~\ref{calib}, we introduce the quasar
sample, and we will argue that it can be considered a representative sample of
the overall quasar distribution. Section~\ref{PhotometricCalibration} goes
through the careful calibration steps needed before one can properly start
interpreting the results. The method outlined is in principle the same as in
Paper~I, but since the sample is much larger, it does allow for some enhanced
corrections. In \S~\ref{IntroStructureFunction}, we will introduce the
variability diagnostic used throughout the paper: the Structure Function
(hereafter SF). This measure has been used extensively in the literature, and
allows for easy and direct comparison with long-term variability studies based
on the monitoring of individual quasars \citep[e.g.,][]{hawkins02}. In addition,
\citet[][]{kawaguchi98} modeled SF behavior depending on the intrinsic
variability mechanism. Clear differences in the SF curves are expected depending
on whether the dominant variability is due to either bursts of supernovae close
to the nucleus, instabilities in the accretion disk, or intervening
micro-lensing events.  Throughout the paper we will refer back to the
predictions made in \citet{kawaguchi98}.

Section~\ref{StellarStructureFunction} describes the results of the calibration
on the stellar SF. The purpose of this detailed section is twofold: first, it
reflects the level of data-quality we have attained with the calibration method,
and secondly, it identifies subtle effects on the data that may have gone
unnoticed by just focusing on the quasar SF. Among other things, the clear
differences in photometric data-quality between the POSS and GSC2 surveys only
shows up significantly in the stellar SF. Also, the Malmquist bias signal is
clearly seen in the stellar SF, whereas it is masked (and indeed washed out) in
the quasar SF by the light-curve asymmetry signal (cf. \S~\ref{LCasymmetry}).

Section~\ref{QSF} and its subsections detail the results we obtained for our
quasar sample. These are discussed in light of the existing literature data and
our last section on SF modeling in \S~\ref{Modeling}.

\section{Sample Selection and Calibration}\label{calib}


\begin{figure}[t]
\epsscale{0.95}
\plotone{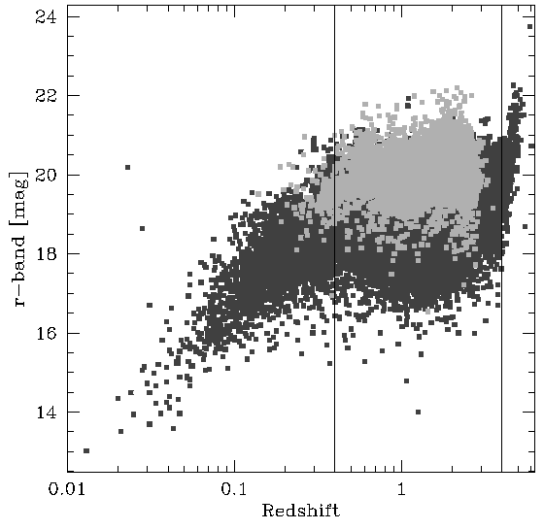}
\caption{Distribution of quasar \rsdss-band magnitudes as function of redshift. The
dark squares are quasars from the DR2 data-set, and the light squares are
(spectroscopically confirmed) 2dF quasars that have photometric DR2 data. The
two vertical lines demarcate the redshift range of 0.4 to 4 for which the
\rsdss-band distribution is more or less independent of redshift. Outside these
boundaries a strong correlation between redshift and magnitude exists.}
\label{brightDist}
\end{figure}

In order to significantly improve on our work in Paper~I (using 3791 quasars
from the SDSS Early Data Release), we had to wait until the later releases would
increase the quasar sample by a large amount. This was accomplished by the two
subsequent Data Releases, which expanded the database first to 16\,908, and then
to 35\,165 quasars (Data Release 2). This DR2 is described in detail in
\citet{abazajian04}. In addition, we added all the 2dF quasars \citep{croom04}
that are covered by the DR2, but are not among the 35\,165 in their quasar
database. This increases the sample size to 41\,578 quasars, all of which have
accurate and recent SDSS photometric information. However, since we are
interested in historic variability, we had to remove the 187 quasars that were
not included in either the POSS or the GSC2 catalogs. This leaves us with the
final sample of 41\,391 quasars. Table~1 has the exact break-down of photometric
information on this sample.

\begin{table}[t]
\caption{Quasar Sample}
\vspace{2mm}
\label{sampletable}
\begin{tabular}{rl}
\hline
\hline
Quasars & Photometric Epochs\\
\hline
32,832 & SDSS\quad GSC\quad POSS~I\\
 8,530 & SDSS\quad GSC\\
    29 & SDSS\quad \phantom{GSC}\quad POSS~I\\
\hline
41,391\\
\hline
\end{tabular}
\end{table}

\subsection{Photometric Properties}


\begin{figure}[t]
\epsscale{1.0} 
\plotone{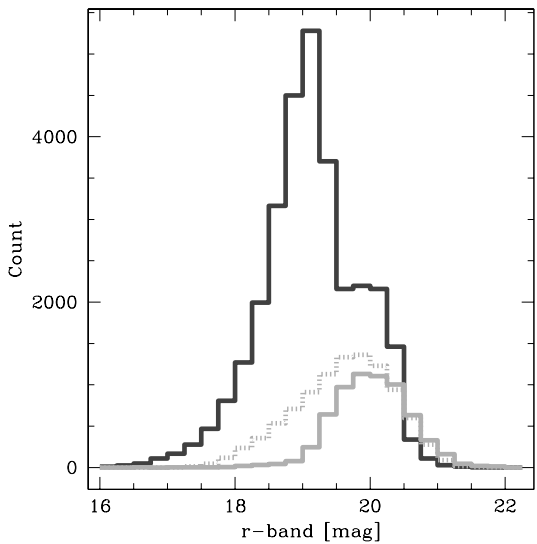}
\caption{Histograms of the magnitude distribution for quasars with redshifts
between 0.4 and 4. The dark-colored histogram is for the DR2 data only, and the
light-colored lines are for the additional 2dF quasars. The dashed histogram
shows all the 2dF quasars in the DR2 area, and the solid gray histogram just the
ones that are not in the DR2 sample. It is clear that different selection
criteria were used for the SDSS and 2dF surveys. The SDSS appears to have a
bimodal magnitude distribution, whereas the 2dF is more uniformly distributed
around \rsdss=19.8.}
\label{brightHist}
\end{figure}

\begin{figure}[t]
\plotone{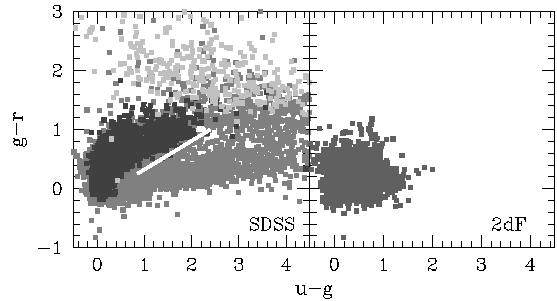}
\caption{Plot of the DR2 and 2dF quasars (left and right panel) in the 
$(u-g)$ and $(g-r)$ color plane. This is one of the planes used to select quasar
candidates for spectroscopic follow-up with SDSS (e.g., Richards et al.  2002).
The stellar locus is indicated by the thick solid white line, and illustrates
the intrinsic color differences between stars and quasars (hence the relative
paucity of quasars in that area). The SDSS sample has been divided into three
redshift bins (colored from dark to light): $z < 0.4$, $0.4 < z < 4$, and $z >
4$.  Note the limited color range of the 2dF quasars compared to the SDSS
selection criteria.}
\label{sampleDiff}
\end{figure}

Figure~\ref{brightDist} shows the \rsdss-band magnitude of the sample as
function of redshift. The faintest quasars are about \rsdss=21, except the very
highest redshift quasars ($z >4$), which clearly have been selected using
different criteria. The lower redshift sources ($z < 0.4$) also show a
correlation between their redshift and optical magnitude. At these redshifts the
quasar host galaxy is contributing significantly to the overall luminosity, and
progressively more so with decreasing redshift. This non-variable host galaxy
component will lower the relative variability amplitude of the AGN. If, for
instance, the AGN varies intrinsically by 10\%, placing it inside a galaxy with
the same magnitude will lower the variability of the combined system to 5\%.
The optical variability of an individual quasar is not just a function of AGN
luminosity relative to its host galaxy luminosity, it also depends on the
redshift of the source.  First, the contrast between the AGN and its host galaxy
increases dramatically toward the restframe blue and UV wavelengths (as probed
by the passbands even at moderate redshifts). Second, the $(1+z)^4$ cosmological
surface brightness dimming factor affects the extended galaxy more than the
point-source AGN contribution, again increasing the contrast between the two
components. Both these redshift dependent trends diminish the unwanted
variability-lowering effect by the host galaxy, and based on
Fig~\ref{brightDist}, it does not appear to contribute beyond $z \approx 0.4$.

The bulk of our quasars (36\, 802) have redshifts between 0.4 and 4.0 (marked in
Fig.~\ref{brightDist}), whereas 4424 (or about 10\% of the sample) are at
redshifts below 0.4 and might potentially be affected by their host galaxies.  A
direct comparison between the results with and without the low redshift
data-set did not yield any significant differences, except at the longest
time-lags in the \rsdss-band in particular (see \S~\ref{galcont}).

The quasars added from the 2dF survey have a different brightness distribution
(cf. Fig.~\ref{brightDist}, light-colored points) than the SDSS quasars, mainly
because of different selection criteria. This difference remains, even if we put
the 2dF and DR2 overlap quasars back into the 2dF sample, as is clearly
illustrated in Fig.~\ref{brightHist}. Both the SDSS and 2dF use multi-band
photometric criteria to preselect for quasar candidates. In addition, the SDSS
sample has been augmented by targeting FIRST and ROSAT counterparts as well
\citep{richards02}, and in general uses a less restrictive color cut.  The
fraction of radio- and X-ray loud sources is relatively low (2692 FIRST
counterparts within 2\arcsec, and 479 ROSAT counterparts within 15\arcsec), so
it does not significantly alter the distribution.

Figure~\ref{sampleDiff} illustrates the differences between the SDSS and 2dF
selected quasars. The sources have been plotted on the $(u-g)$ and $(g-r)$
plane, which is one of the color-color diagrams used in selecting SDSS quasar
candidates \citep{richards02}. The 2dF quasar candidates were selected from
scanned UK Schmidt Telescope (UKST) photographic plates, with magnitudes ranging
from $18.25 < b_{\mbox{J}} < 20.85$. In addition, the candidates had to satisfy
one of the following criteria \citep[see][]{croom04}: $u-b_{\mbox{J}} \leq
-0.36$; $u-b_{\mbox{J}} < 0.12 - 0.8(b_{\mbox{J}}-r)$; or $b_{\mbox{J}}-r <
0.05$. This results in a markedly different color and ($r$-band) magnitude
distribution from the SDSS quasars. However, the important similarity is that
the redshift distribution between $0.4 < z < 4$ is fairly uniform as function of
$r$-band magnitude (cf. Figs.~\ref{brightDist} and \ref{brightHist}). So, even
though the quasars have been selected differently, and actually populate the
color-color diagram of Fig.~\ref{sampleDiff} differently, we feel that there is
no a-priori bias in either sample with respect to variability in general, and
variability on select time-scales in particular.

\subsection{Photometric Calibration}
\label{PhotometricCalibration}

Calibration of historic photographic plate material can be achieved by virtue of
using large numbers of random field stars around the (quasar) position of
interest. Plate-to-plate variations in emulsion quality, and even variations
within a single plate can contribute significantly to measurement uncertainties.
So, even though the POSS~I and GSC2 catalogs have been calibrated carefully (as
a whole), and brought up to CCD photometric standards, there is still a lot of
improvement to be made by recalibrating the photometry. We basically follow the
same procedure as outlined in Paper~I by using all the available photometry for
the field stars within 5\arcmin\ of the quasar position. This typically amounts
to (depending on the epoch) anywhere between 50 to 500 stars. We like to stress
that this ``local'' calibration is to be preferred over complete plate
corrections due to the potential inhomogeneities inherent to photographic plates
\citep[e.g.,][]{lattanzi91,gal03}.

We will go over the calibration process step by step, but we will refer to the
calibration sections in Paper~I where appropriate. Most of the next discussion
will highlight the improvements we were able to make on the old procedure,
mainly due to the much larger data-set.

The first step is to calculate the best passband transformations for each quasar
individually, using the nearby field stars. The transformations involved are,
for the POSS~I: $B$ to \gsdss, and $R$ to \rsdss. Note that the $B$ and $R$
magnitudes are already transformed to the Johnson passbands from their
photographic $O$ and $E$ emulsions (see Reid et al. 1991, and Monet et al. 2003
for the $B$ and $R$ transformations). For the GSC2 plates, the relevant color
transformations are $J$ to \gsdss, and $F$ to \rsdss. In principle, our
transformation will take care of the proper passband corrections, possible plate
/ weather variations, and the fact that SDSS uses AB magnitudes whereas the
catalogs are on the Vega system.  However, an important caveat we like to
emphasize here is that the ``best transformation'' is defined as the particular
transformation that results in the smallest color rms {\it for the stars}. As we
have explained in Paper~I, this does not necessarily translate into the best
calibration for the quasars, which after all, is the transformation we are
interested in. There are two main contributors to this stellar-quasar disparity:
their optical spectrum is completely different (cf.  Fig.~3 of Paper~I), and
quasars typically have powerful emission lines which depending on their
redshift, may, or may not, be present in the passband.  These emission lines can
account for upward of a few tenths of a magnitude of the total brightness. Both
these differences between the field stars and the quasars render the stellar
transformation less than ideal. It is something we can correct for, however.


\begin{figure}[t]
\epsscale{1.0}
\plotone{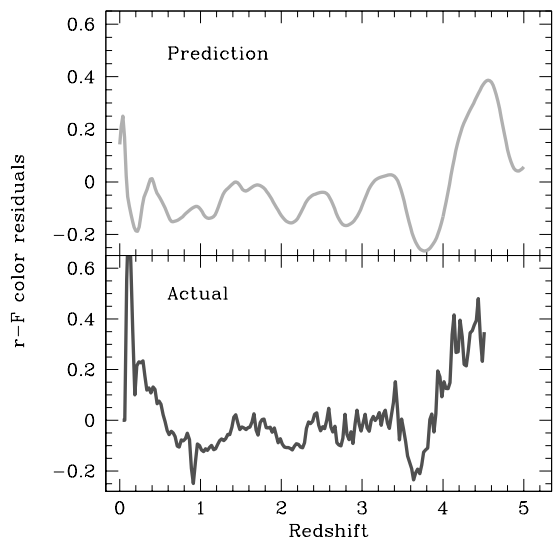}
\caption{Residual \rsdss$-F$ color differences after applying the best stellar
transformation to the quasar magnitudes, as function of redshift. The lower
panel shows the actual median of the distribution ($\sim 40\,000$ quasars). The
top panel depicts the expected color changes, based on a quasar template
spectrum, and a mean stellar spectrum of a K2{\small V} star (cf. Paper~I). Note
the excellent agreement between the two curves, except for the lowest redshift
range (which is affected by the host galaxy contribution, cf.
\S~\ref{galcont}). }
\label{compTemplate}
\end{figure}

Figure~\ref{compTemplate} illustrates this calibration best. The bottom panel
shows the median \rsdss$-F$ color for our quasar sample as function of redshift.
Ideally, these residuals should be close to zero after calibration in the
absence of emission lines and quasar-stellar spectral differences. This is
clearly not the case. However, these color excursions (of up to more than 0.2 in
magnitude) are closely matched by what one would expect using quasar and stellar
template spectra (top panel). As explained in Paper~I, we get the best agreement
between the actual residuals and the theoretical ones by assuming a mean stellar
template of a K2{\small V} star. This stellar type is consistent with
expectations based on population models for our Galaxy
\citep[e.g.,][]{bahcall80, bahcall81}.  So, the fact that these color excursions
are well understood in terms of quasar emission lines moving in and out of the
observing passband, makes it clear that we have to correct for it. If left
``untreated'' it will affect the variability SF directly by artificially
inflating the rms values at certain time-lags. Given the epoch distribution of
the observations (time-lags preferentially at $\sim1$, $\sim10$, and $\sim50$
years), the redshift maps more or less directly onto a particular time-lag. The
strong excursion at $z\approx 3.6$, for example, would skew the SF signal
preferentially at $10 / (1+3.6) \approx 2$, and $50 / 4.6 \approx 11$ years.  In
this paper we opted to use the actual median, as calculated across bins with a
width of 0.05 in redshift units, over the modeled offsets. This accounts much
better for the low ($z<0.4$) redshift quasars which are increasingly more
contaminated (with decreasing redshifts) by their host galaxies.

All of the other passband transformations are treated similarly. The result is
that each historic passband has been brought onto their SDSS counterpart (either
$g$ or $r$), with the important distinction that the color distributions are
centered around $0$ as a function of redshift. This method improves
significantly over the procedure outlined in Paper~I. There, bulk corrections
have been applied to the color distributions (irrespective of redshift).
Figure~2 of Paper~I can therefore be considered a projection of
Fig.~\ref{compTemplate} onto the $y$-axis. Only with the large increase in
sample size were we able to actually correct for the redshift dependence in a
meaningful way.


\begin{figure}[t]
\epsscale{1.0}
\plotone{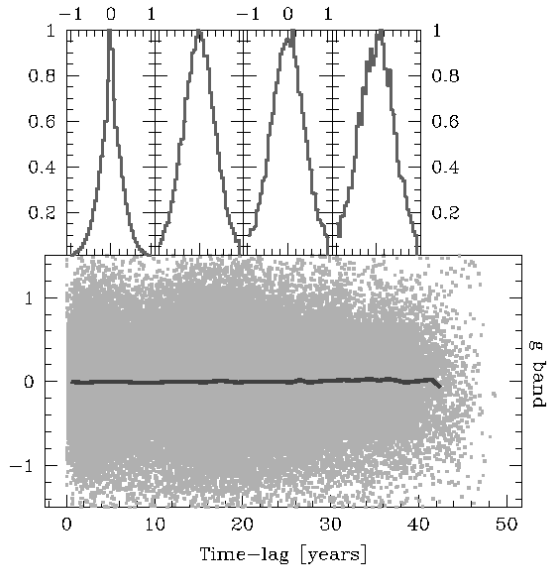}
\caption{Quasar variability distribution in the \gsdss-band as function of time-lag.
This distribution has been calibrated as described in the text. The thick solid
line indicates the local median value of the distribution, and its lack of
significant deviation from zero serves as an indication of our careful
calibration. The top 4 histograms are for the time-lag bins [0, 10$>$, [10,
20$>$, [20, 30$>$, and [30, 40$>$ years respectively. The increase of the FWHM
with increasing time-lag is evident. The bin values are: 0.57, 0.94, 1.00, and
1.05 magnitude. Note that the first histogram deviates by quite a bit from a
Gaussian distribution. The other three are accurately described by one.}
\label{permCheck}
\end{figure}

After all the photometric data have been transformed onto the SDSS passbands,
the measurements for each individual quasar are permutated among each other,
resulting in about 4 time-lag measurements per band per quasar.  Obviously, none
of the individual quasars have been sampled photometrically anywhere near enough
to produce a meaningful structure function for each quasar individually. The
combined data-set, however, allows for detailed variability studies provided one
assumes that the underlying cause of quasar variability is the same for all of
them. We will get back to this issue in \S~\ref{Modeling}.

This final data-set (one each for the \gsdss- and \rsdss-bands) contains sorted
pairs of time-lag (in years) and magnitude difference. For our sample of 41\,391
quasars, this amounts to 170\,102 individual measurements for the \gsdss-band,
and 131\,123 for the \rsdss-band. This exceeds the total number of permutations
in Paper~I by more than an order of magnitude.  The difference in the totals
between the bands is because some GSC2 photometry for the quasars have been
repeated more in the $J$ than in the $F$ band, boosting the permutation numbers
for the \gsdss-band.

The actual data for the \gsdss-band have been plotted in Fig.~\ref{permCheck}.
The bottom panel shows the magnitude differences as function of intrinsic
time-lag. Since the quasars are quite spread out in redshift space (cf.
Fig.\ref{brightDist}), we have to bring the actual time separation between the
observations onto the reference frame of the quasar itself (by dividing it by a
$(1+z)$ factor). This has the additional advantage of smoothing out the time-lag
distribution. So even though the observing campaigns were well separated in time
(1950's, 1990's, and $\sim 2000$), resulting in time-lags clustering around a
few, $\sim 10$, and $\sim 50$ years, the $(1+z)$ redistributing factor results
in a pretty smooth distribution up to time-lags of $\sim 40$ years (cf.
Fig.~\ref{permCheck}, bottom panel).

The top four panels of Fig.~\ref{permCheck} provide a direct picture of the
increase in FWHM (and hence the rms) of the magnitude difference distribution as
time-lags increase. This is actually the definition of the SF (see next
section). Since the total number of time-lag measurements decreases with
increasing time-lag, this is not immediately obvious looking at the point-cloud
in the lower panel. The numbers of data-points for the current 10-year time-lag
binning are: 86\,795, 50\,974, 22\,627, and 8\,520 permutations respectively.
While the numbers do decline, they are still large enough to assess the FWHM of
the distribution very accurately. The last bin alone already contains 30\%\ of
the total number of permutations used for Paper~I.

\section{Structure Function} 
\label{IntroStructureFunction}

Our analysis of Paper~I, and the current paper, will utilize the SF as the tool
to characterize the quasar variability.  SF's are not very sensitive to aliasing
problems due to discrete and/or sparse time sampling \citep[e.g.,][]{hughes92},
which make them well suited for our purpose. As before, we define the SF as:

\begin{equation}
S(\tau) = \left( \frac{1}{N(\tau)} \sum_{i<j}{[m(i)-m(j)]^2} \right)^{\frac{1}{2}}
\end{equation}

\noindent with the summation over all the combinations of measurements for which 
$\tau = t_j-t_i$. In our case we group all the $n(n-1)/2$ permutations into bins
which contain at least 200 measurements. The SF value for each bin is then given
by the rms of the magnitude permutations. 

\subsection{Error estimates}\label{errorest}

This results in $\sim$1500 bins, which are then binned again onto a fixed grid
in log time-lag space (running from $-0.97$ to 1.55 in 0.06 dex bins for a total
of 42). This facilitates easy comparison between model and actual SF curves. It
also allows us to approximate the error on a particular SF point by calculating
the rms of the $1500/42 \approx 36$ values inside each bin. This is basically
the same method as we employed in Paper~I. The presented error-bars reflect
therefore accurately the actual local SF uncertainties. It should be stressed
that, unlike a well monitored SF of a {\it single} source for which all of the
bins are cross-correlated with each other and an objective error estimate is
hard to give, our bins are essentially {\it independent}. Out of the 150\,000 or
so time-lag measurements (per band) only measurements for a single quasar (about
4) are correlated with each other. In other words, each of the SF bins contains
a virtually completely different set of quasars. This bin-independence also
allows us to quantify SF similarities in terms of their offset distributions.
Assuming two SF curves, labeled $A$ and $B$, both of which are binned to the
same $N=42$ bins specified above, we can define:

\begin{equation}\label{oline1}
\overline{O} =  \frac{1}{N} \sum_{i}^{N}{S_A(i)-S_B(i)}
\end{equation}
\begin{equation}\label{oline2}
\Delta\overline{O} =  \frac{\sigma}{\sqrt{N}} = \frac{1}{N} \left( 
\sum_{i}^{N}{\left(S_A(i)-S_B(i)-\overline{O}\right)^2} \right)^{\frac{1}{2}}
\end{equation}

\noindent after substituting $N\approx \sqrt{N}\sqrt{N-1}$. The quantities
$\overline{O}$ and $\Delta\overline{O}$ represent the mean SF offset and its
$1\sigma$ uncertainty, respectively. We will use this metric in particular for
our SF asymmetry part of the paper.

\subsection{Stellar Structure Function}
\label{StellarStructureFunction}

The SF for the calibration stars serves multiple purposes.  If we assume that
stars, {\it on average}, are not variable, then the SF derived from it should
not exhibit any correlation with time-lag. In other words, it should be parallel
to the $x$-axis (in plots like Fig.~\ref{SFstars}). This was indeed found to
be the case for the stars in Paper~I, which clearly illustrated the significant
differences between the SF behavior of stars and quasars. However, given the
much smaller sample sizes for Paper~I (note that the number of calibration stars
is linked to the number of quasars), the overall stellar SF was rather noisy. It
just served to make the point that constructing an SF from a random sample of
stars resulted in a non-variable SF curve, but it clearly was not good enough to
go beyond that. The current sample, however, is large enough. In the next few
sections, we will discuss the stellar SF in more detail.

\subsubsection{Stellar Type Dependencies}
\label{stellartypedep}

In the same way spectral differences between the average stellar spectrum and a
quasar spectrum lead to slightly different passband corrections, and therefore,
additional noise to the variability measure, spectral differences among stars
themselves will inflate its SF variability signal as well. This has to be
considered in the construction of the stellar SF.  The reason we can use stars
to calibrate the {\it quasars} at all, is that the {\it mean} of the stellar
color distribution does not change that much going from one sightline to
another. The stellar population therefore does not change a lot across the sky
covered by DR2\footnote{It should be noted in this respect that the DR2 does not
cover the galactic plane.}.

\begin{figure}[t]
\epsscale{1.0} 
\plotone{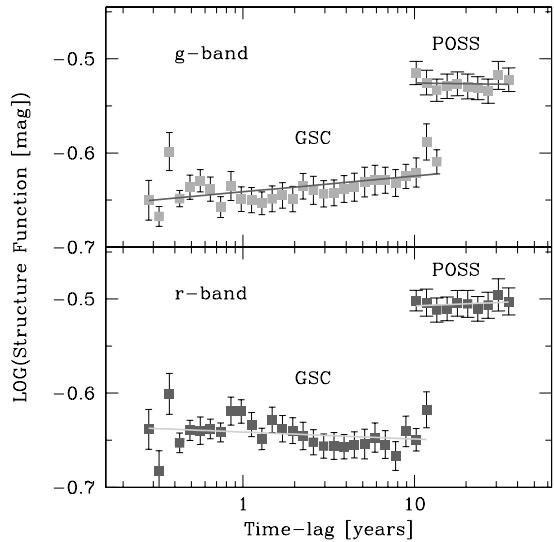}
\caption{Structure functions of calibration stars; SDSS $g$-band in the top panel,
and SDSS $r$-band in the bottom panel. The data quality of the earlier POSS
survey is lower than that of the GSC, resulting in a slightly higher noise
plateau. The solid lines are least squares fits to the data points. The slopes
are: $+0.017\pm0.006$ (GSC-$g$), $-0.003\pm0.012$ (POSS-$g$), $-0.008\pm0.006$
(GSC-$r$), and $+0.007\pm0.009$ (POSS-$r$).}
\label{SFstars}
\end{figure}

In order to limit the stellar spectral range allowed for our template SF, we
only included stars within a magnitude range ($17 < r < 21$), and an $(r-g)$
color within 0.2 magnitudes of the typical stellar color of $(r-g)=0.4$
\citep[cf.][]{stoughton02}.  This color cut effectively limits the allowed range
of stellar colors, and improves the passband calibrations accordingly. The
resulting time-lag permutation database contains 2.1 million data-points (over
both bands), an order of magnitude larger than the quasar permutation database.
The net effect is a lowering of the SF, especially for the GSC data (below
time-lags of 10 years).  The POSS~I data, plotted separately in
Fig.~\ref{SFstars}, retains a slightly higher noise plateau, mainly due to the
photometric data quality differences between the GSC and POSS surveys.


Because of this data quality difference, we have separated out the GSC and POSS
contributions to the stellar SF in Fig.~\ref{SFstars}. Linear least squares fits
to the data-points have been made, and their slopes are: $+0.017\pm0.006$
(GSC-$g$), $-0.003\pm0.012$ (POSS-$g$), $-0.008\pm0.006$ (GSC-$r$), and
$+0.007\pm0.009$ (POSS-$r$). With the possible exception of the $g$-band SF for
the GSC data, none of the slopes differ significantly from zero (i.e., no
correlation with time-lag). Even the GSC-$g$ case is only weakly increasing with
time-lag with the highest SF value within $1\sigma$ of the lowest. It should
also be noted in this respect that the small SF excursions from the mean in
the GSC-$r$ case are not statistically significant. 

The main result from this exercise is twofold. First, there is no significant
correlation between the SF for stars and the time-lag for both the GSC and POSS
data.  This implies that any signal we detect in the {\it quasars} beyond the
stellar SF curve must be due to the quasars themselves.  Also, it is not clear
how much of the stellar SF signal is due to intrinsic color scatter, and how
much can be attributed to measurement noise. Recently, \citet{sesar04} estimated
the photometric error in the GSC2 and POSS~I to be 0.10 and 0.15 magnitude,
respectively. They used individual SDSS plates (100 square arcminutes) to
correct the POSS~I and GSC2 photometric catalogs.  Taken at face value, these
uncertainties translate into a white noise signal in the SF at levels of $-0.85$
and $-0.67$, using Eqn.~3 from Paper~I.  Both these levels are lower than the
plateaus we measure in our stellar SF.  How much of this difference can be
attributed to the different calibration method, and how much can still be
improved upon by even more carefully designed color-cuts, is unknown. As a
consequence, we cannot ``correct'' the quasar SF by subtracting a
measurement-noise component, and instead, like in Paper~I, we will have to
include a white noise term in our Monte Carlo models.

Figure~\ref{SFstars} presents the final corrected SF for both the \gsdss- and
\rsdss-bands. As measured from the figure, the final SF levels, separated into
time-lags less and more than 10 years, are: $-0.63$ and $-0.53$ for \gsdss, and
$-0.64$ and $-0.51$ for \rsdss. The corresponding formal noise levels are: 0.17,
0.21 and 0.16, 0.22 magnitudes. The POSS~I levels are only slightly lower than
found in Paper~I, but the GSC2 levels are suppressed by about 0.10 (in SF
units).

\subsubsection{Malmquist Bias in the Structure Function}
\label{malmquistSF}


\begin{figure}[t]
\epsscale{1.0} 
\plotone{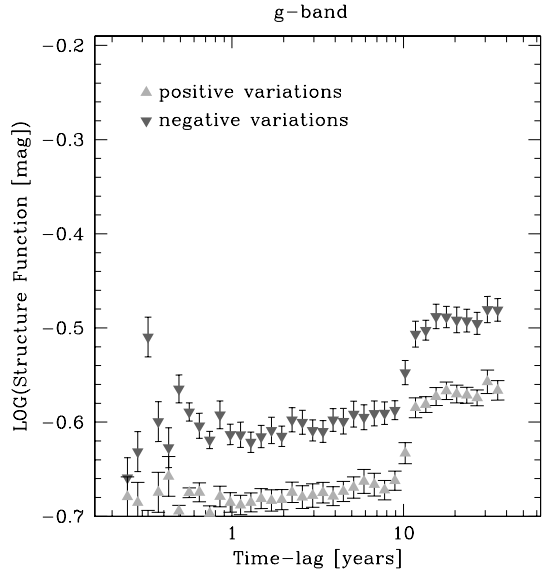}
\caption{The $g$-band Structure Functions for the calibration stars (using the narrow 
$(r-g)$ color set), separated in positive-only (light-gray symbols) and
negative-only (dark-gray symbols) variations. Positive variations are defined as
brightening with increasing time (i.e., the SDSS epoch is the brightest), and
negative variations have the opposite sign (i.e., the SDSS brightness is less
than what it was at the older epochs).  The {\it offset} between the two SF
curves is due to the Malmquist bias, acting upon the subset of the calibration
stars that are variable. Note that, unlike in Fig.~\ref{SFstars}, the GCS and
POSS data have been combined, resulting in a slightly smoother transition across
time-lags of 10 years.}
\label{SFposneg}
\end{figure}

Leading up to the construction of the SF is an intermediate step in which we
take care to center the distribution of variations around $0$, as function of
time-lag (see Fig.~\ref{permCheck} for the quasar case), without changing the
{\it shape} of the distribution. If there is any asymmetry present in the
distribution, we can test for it by looking at the positive and negative sides
of the distribution separately.  The variations are defined with respect to the
newest epoch, so positive variations imply that the source is brighter (i.e.,
had a lower magnitude) than it was before, and negative variations imply that
the source is fading with time. This definition is consistent with the one used
in \citet{kawaguchi98}.

For a perfectly symmetric variation distribution, the positive and negative SF
curves should be identical. This, however, is clearly not the case for our
calibration stars. Figure~\ref{SFposneg} shows the SF for the positive
(light-gray triangles) and negative variations (dark-gray triangles) separately.
Since the SF curves are not identical, it indicates an asymmetric variation
distribution. In the next few paragraphs we will argue that this asymmetry is
due to the Malmquist bias in our stellar sample. It basically means that for the
subset of our stars that have intrinsic variability, the ones that were a lot
fainter at the earlier epochs (i.e., the positive variations) would not make it
into either the POSS~I or GSC2 catalogs (given their brightness limits), thereby
reducing the rms / SF signal.  Variations in the other direction are not
affected by this since there is no upper brightness limit to the catalogs.  The
net effect is a skewing in the variation distribution.

There are some important points to be made based on the SF curves in
Fig.~\ref{SFposneg}. First, both SF curves are the same, except for a constant
offset in log (which translates into a $\sim14\%$ increase in rms). This implies
two things: 1) The Malmquist bias does not depend on time-lag (and it should
not), nor does it depend on the quality of the photometry. The POSS~I
measurements are noisier than the GSC2 ones, but there is no evidence for a
different offset between the positive and negative SF curves.  2) The magnitude
of the stellar variability is not correlated to intrinsic time-lag because we
are not sensitive to their time-scales, quite unlike the variations for the
quasars (see \S~\ref{QSF}).

The second observation we like to make is that, on top of the uncertainty about
the exact contribution of measurement noise to the SF (cf.
\S~\ref{stellartypedep}), this asymmetry induced SF signal further compounds the
problem of disentangling the SF into its various contributions. Therefore, we
cannot use the stellar SF to improve or correct the quasar SF.

\subsubsection{Variable star contribution}

\citet{sesar04} quote a variable star fraction (with variability exceeding 0.2
magnitude) of at least 1\%\ of the population\footnote{Of the population away
from the Galactic plane.}. The rms difference between the positive and negative
SF curves (from Fig.~\ref{SFposneg}) is actually not that large: for the shorter
timescales, we measure rms = 0.148 and rms = 0.177 mag for the positive and
negative SF curves. Since this 0.029 magnitude difference is not that large, a
small fraction of variable stars might be enough to skew the distribution. For
sufficiently large distributions (where $(N-1)/N \approx 1$), the rms of two
distributions, each with its own $\sigma$, can be combined as follows:

\begin{equation} \label{rmses}
rms = \left( \frac{N_1}{N_1+N_2} \sigma_1 + \frac{N_2}{N_1+N_2} \sigma_2 \right)^{1/2}
\end{equation}

\noindent with $N_1$, $N_2$ the total number of items in each distribution. In our
case, we assume the following for the {\it non-variable} part of the stellar
distribution: $\sigma_1 = 0.13$, and $N_1 = 99 N_2$ (i.e., only 1\%\ of the
sample is variable). We also have to assume that this $\sigma_1$ scatter is
symmetric around $0$, and is small enough not to be affected by the Malmquist
bias. It is in effect a constant contribution to both of the SF curves. All of
the Malmquist signal therefore has to be ascribed to the variable subset of
stars.

We can make the observations agree with our simple distribution model by
assuming that half of the variable stars vary with on average 0.5 magnitude, and
the other half varies by 1.5 magnitude. Since we are looking out of the Galactic
plane, and do not make any a-priori assumptions about the stars (other than that
they are found close to a quasar), the main constituents of the variable star
population are RR-Lyrae and Asymptotic Giant Branch (AGB) stars. Neither the
magnitude of their variation, nor their relative fractions are inconsistent with
the values we assume here \citep[see, e.g.,][]{derue02}.

It is the high variability part that drops out on the positive variation side of
the distribution (i.e., the half with the 1.5 magnitude variability).  This
results in the following rms values: $\sigma_{\mbox{neg}} = 1.12$ and
$\sigma_{\mbox{pos}} = 0.50$ magnitude.  By applying Eqn.~\ref{rmses}, we arrive
at total rms values of 0.148 and 0.179, very close to the actual values.
Obviously, none of these values are constrained to any degree, they just serve
to demonstrate that we can actually explain the observed stellar SF curves by
the Malmquist bias due to a small percentage of variable stars.

We will come back to the asymmetry issue in the next Section, where we discuss
the SF of our quasar sample.

\subsection{Quasar Structure Function} 
\label{QSF}


\begin{figure}[t]
\epsscale{1.0} 
\plotone{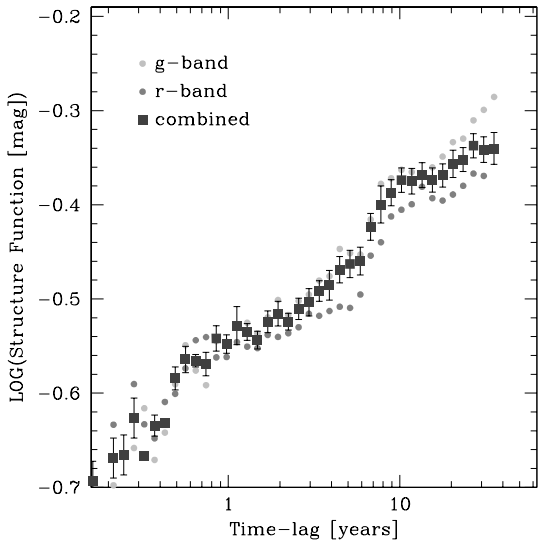}
\caption{Combined \gsdss- and \rsdss-band structure function for the quasar
sample (dark squares). The least-squares slope of the SF is: $(0.153\pm0.004)$
with intercept $(-0.557\pm0.003)$ at time-lags of 1 year. The SF curves for the
\rsdss- and \gsdss-bands are indicated separately. Note that we limited the
redshift range for the \rsdss-band curve to $z > 0.6$ (see
\S~\ref{galcont}). The clear break in the stellar SF curve at low noise levels
(cf. Fig.~\ref{SFstars}), due to the varying data quality, does not affect the
SF curve for quasars.}
\label{SFquasars}
\end{figure}

Our quasar SF curve is presented in Fig.~\ref{SFquasars} by the dark-gray
squares.  It is immediately clear that this SF curve is a vast improvement over
the one presented in Paper~I (Fig.~7), with a much smaller scatter of the
points.  The individual error-bars are actually small enough to allow for
detailed modeling, something that the data quality did not allow for in Paper~I.
Before we continue, however, we like to establish the reality of the quasar
SF curve by comparing it to currently the best long-term one available in the
literature.

\subsubsection{Literature comparison}


\begin{figure}[t]
\epsscale{1.0} 
\plotone{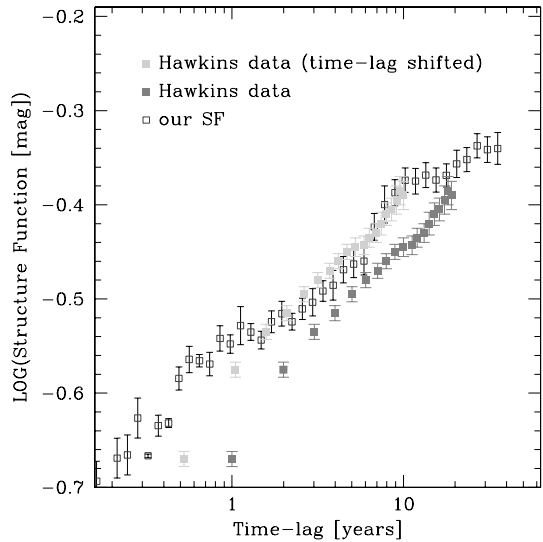}
\caption{Comparison of our quasar SF with a literature SF based on
long-term monitoring of 401 quasars (Hawkins 2002). The original Hawkins data
have not been corrected for time-dilation (dark-gray squares), but can be made
to agree with our SF by applying a 0.3 dex shift toward shorter time-lags
(light-gray squares). This corresponds to a sample mean offset of $z=1$. Note
that, except for the 2 shortest time-lag bins, all the Hawkins data points are
within $1\sigma$ of our points. Our short time-scale measurements are most
likely shifted upward a bit due to the intrinsically higher level of measurement
noise (compared to the Hawkins data set).}
\label{SFhawkins}
\end{figure}

This paper's approach to long-term quasar variability is quite different from
the monitoring approach, in which one repeatedly measures the brightnesses of a
fixed sample of quasars. Given our statistical method, which uses archival
photometry data and a less than straightforward calibration path, one might have
concerns about the results. We therefore compare our resulting quasar SF to the
highest quality available long-term quasar SF from the literature (taken from
Hawkins 2002). This SF was constructed based on $\sim30$ year monitoring data on
a sample of 401 quasars, and is plotted with our SF in Fig.~\ref{SFhawkins}.
The Hawkins SF, which has not been corrected for time-dilation, is represented
by the dark-gray squares. Since our data have been corrected by the $(1+z)$
term, we have to shift the Hawkins SF over toward shorter time-scales. The
light-gray squares have been shifted leftward by 0.3 dex (corresponding to a
sample mean redshift of 1). This shifting does not affect the slope of the SF,
as mentioned in, e.g., \citet{kawaguchi98, hawkins02}. Note that, with the
exception of the shortest two time-lag data points, all of the Hawkins data can
be made to fall within $1\sigma$ of our curve\footnote{By applying a reasonable
time-dilation correction. We do not have access to their quasar redshifts.}.
The offset at short time-scales is most likely due to the higher photometric
accuracy (and hence a correspondingly lower noise plateau) of the Hawkins data
compared to our data (for which these data points will be shifted upward a bit,
see Fig.~6 from Paper~I). 

It is reassuring to see that these two different approaches, each with their
own set of clearly distinct potential systematic problems, produce SF curves
that are so alike. 

We do like to remind our reader that, even though the curves are similar, there
is a key difference: our data bins are almost completely independent with
different quasars contributing to different bins, whereas in the Hawkins SF,
most bins contain data (permutations) from the same set of quasars. This is a
big advantage when one tries to understand the SF errors (cf. \S~\ref{errorest})
and the potential differences between positive- and negative-variation SF curves
(cf. \S~\ref{LCasymmetry}).

\subsubsection{Lack of SF turnover}

There are a few things we like to discuss based on our quasar SF in
Fig.~\ref{SFquasars}. First of all, there is no indication that the SF curve is
turning over, consistent with the results from Hawkins (2002). What we are
looking for in the SF curve is the presence of a consistent plateau beyond a
certain time-scale (cf. Fig.~\ref{SFqmodel}, and \citet{hughes92}), and not
necessarily an actual ``peak'' in the SF curve. A significant drop in the SF
signal beyond a particular time-scale usually indicates problems with adequate
time sampling at those time-lags. In our case, the SF curve might be affected
beyond $\sim 40$ years due to the decrease in available time-lags (remember that
the intrinsic time-scale are shortened by the $(1+z)$ time-dilation factor). We
do not see such a drop, however, and the possible leveling off seen in the last
few bins is not significant enough to claim we have detected a preferred
variability time-scale.

This is not to say that there is no upper bound to the variability time-scale,
just that we do not have any sensitivity to it. 

\subsubsection{Color dependencies}

The second clear trend in our SF is that the \gsdss\ SF curve has more signal
(i.e., is more variable at any given time-lag) than the \rsdss\ SF curve. This
is consistent with quasars being intrinsically more variable at shorter
wavelengths. This is not a new result (it was also present, albeit at rather low
significance, in Paper~I), but is now very clearly detected.

\citet{giveon99} measured a $0.02$ magnitude rms difference between B- and
R-band variability in their sample of 42 Palomar Green (PG) quasars.
\citet{trevese02} found variability shifts of these magnitudes between the blue
and red to be consistent over a range of samples. Wavelength variations in the
near-IR tend to be smaller, and are possibly too small to be measured
\citep{enya02a}. Possible mechanisms for these spectral variations include
nuclear star-bursts / supernovae which are predominantly blue
\citep[e.g.,][]{aretxaga97, cidfernandes00}, and instabilities in the nuclear
accretion disk \citep[e.g.,][]{kawaguchi98,giveon99,trevese02}.


\begin{figure}[t]
\epsscale{1.0} 
\plotone{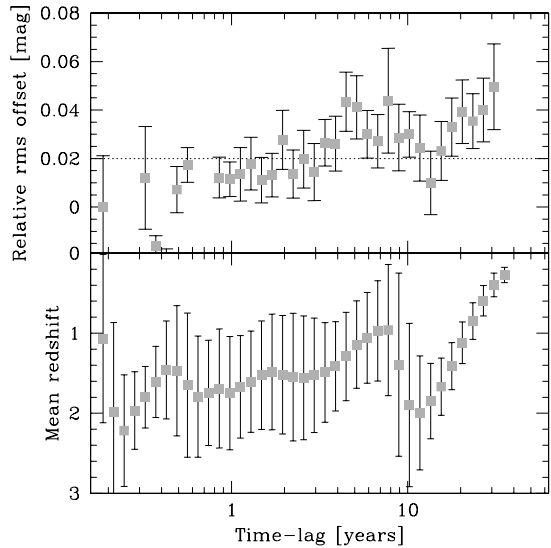}
\caption{The top panel depicts the change in the offset between the \gsdss- and
\rsdss-band variability as function of time-lag. This offset is not constant
since the $r$- and $g$-band SF curves in Fig.~\ref{SFquasars} are diverging.
The dashed line is the constant $0.02$ mag color offset as measured by
\citet{giveon99}.  Our mean value offset value is $0.027$, for time-lags larger
than 1 year (the shorter time-lag data are rather noisy). In the bottom panel,
the mean redshift as a function of time-lag is plotted. The error-bars indicate
the 1$\sigma$ spread in the $z$-distribution within the bin, and should not be
taken as the error in the mean redshift value (which is given by
$\sigma/\sqrt{N}$). The mean redshift and relative color offsets are correlated
to better than 99.9\%\ significance.}
\label{timelagVSz}
\end{figure}

Figure~\ref{timelagVSz} shows the SF color offset as a function of time-lag.  In
the top panel, we have plotted the observed rms offset between the \gsdss- and
\rsdss-band SF curves, as function of time-lag in the quasar restframe. It
appears that the increase in color offset correlates with time-lag, but it is
actually the redshift that the color offset depends on. In the bottom panel we
have plotted the mean redshift of the quasars contributing to a particular
time-lag. Since the time-lags have been converted into the restframe of the
associated quasar, a $(1+z)$ factor has shortened the epoch separation. This
implies that the very longest time-lags ($>40$ years) can only include data from
the lowest redshift quasars, a trend clearly seen in the plot.  Now, assuming
the quasar variability introduces a spectral slope change \citep[i.e., it is
bluer during an outburst, cf.][]{trevese02, vagnetti03} the biggest color
contrast is attained when the observed passbands are furthest apart. This would
be the case for redshift zero objects. At higher redshifts, one starts to probe
progressively bluer parts of the spectrum, and the spectral separation between
the \gsdss- and \rsdss-bands becomes smaller and smaller. This effect of
decreasing color contrast with increasing sample mean redshift is accurately
portrayed in Fig.~\ref{timelagVSz}. Even the sudden increase in mean redshift
around the 10-year time-lag bins is reflected by the drop in the relative rms
change. The Spearman rank coefficient for the correlation between the mean
redshift and the color offset (beyond time-lags of 1 year) is $-0.73$.  This
translates for the 23 degrees of freedom into a less than 0.1\%\ likelihood that
the correlation is by chance.

This good correlation between the two suggests an {\it intrinsic} origin to the
variability. However, this does not rule out the per definition extrinsic
micro-lensing scenario.  It is possible that by assuming a $(1+z)$ time-dilation
correction in the first place, we have introduced some correlation between mean
redshift and color offset.  Given the fact that we rely on the $(1+z)$ term to
smooth out our time-lag coverage, we cannot produce a similar plot {\it without}
such a $(1+z)$ correction to test this.  Nevertheless, a stronger argument
against a lensing scenario is the presence of a light-curve asymmetry signal in
the SF curve (cf.  \S~\ref{LCasymmetry}).

\subsubsection{SF slope}\label{sfslope}

The predicted slopes for the starburst (SB) and accretion disk instability (DI)
models are $(\alpha=0.83\pm0.08)$ and $(\alpha=0.44\pm0.03)$, respectively
\citep{kawaguchi98}. Note that both these slopes are significantly steeper than
our measured value of $(\alpha= 0.153\pm0.004)$, which is far closer to the
modeled micro-lensing slopes ($\alpha\approx0.25\pm0.03$), and a bit shallower
than the $0.20\pm0.01$ slope of the Hawkins (2002) data.

It was this inconsistency between the measured quasar SF slope and the predicted
SB and DI slopes that led Hawkins to propose a micro-lensing origin of long-term
quasar variability. Our data, based on the slope of the SF alone, seems to
support this contention. It also implies that, in the case micro-lensing is
ruled out, that either or both of the other models need significant modification
to explain the observed slope.

As we pointed out in Paper~I (Fig.~6), the measurement noise does have a direct
effect on the slope of the SF. The larger the noise, the shallower the slope.
Since our data is noisier than the photometric observations used by
\citet{kawaguchi98} for their modeling, we will come back to the slope issue in
\S~\ref{intslope} where we construct a noise-less SF.

\subsubsection{Host galaxy contamination}
\label{galcont}


\begin{figure}[t]
\epsscale{1.0} 
\plotone{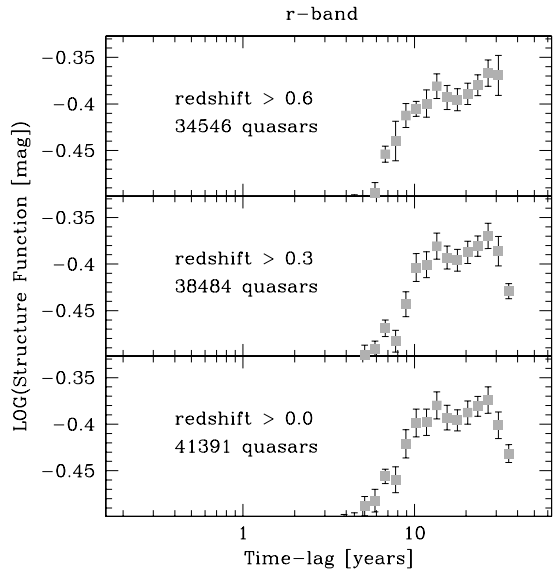}
\caption{Effect of the quasar host galaxy on the AGN variability SF 
(for the \rsdss-band in this case).  The lowest redshift sources contribute the
most to the longest time-lags. It is these bins that are most affected, due to a
combination of cosmological surface brightness dimming and the progressively
smaller (red) galaxy contribution at blue and UV restframe wavelengths.  The
three panels show that the ``turn-over'' at long time-lags disappears if one
increases the low-redshift cut-off. Also none of the other bins are affected, a
nice illustration of the lack of correlation between the bins. Note that we only
detected this effect in the \rsdss-band and not in the \gsdss-band.  This is
consistent with the notion that the host galaxies are intrinsically red.}
\label{SFhostgalcont}
\end{figure}

Variability in quasars is associated with their AGN. A large, non-varying, host
galaxy might therefore limit the relative variation if one uses integrated
magnitudes. Since we are not attempting any galaxy / AGN decomposition, and just
use the total magnitude, this might be somewhat of a concern. The sharp upturn
in residual $r-F$ colors toward low redshifts (cf. Fig.~\ref{compTemplate}, or
Fig.~1 of Paper~I) already hinted that this might be at play at low redshifts.
In this section we will quantify this effect on the SF curve.
Figure~\ref{timelagVSz}, bottom panel, shows that the low-redshift sources
dominate the longer time-lag bins, and that the largest effects are to be
expected here. This is confirmed in Fig.~\ref{SFhostgalcont}, which plots the
long time-lag part of the SF, with different low redshift thresholds. The
apparent turnover at the extreme end of the bottom SF curve (which includes all
data), disappears if one removes all the quasars below redshifts of 0.6.
Evidently, the host galaxy contribution is enough to lower the variability
signal at these redshifts. Clearly, if one wants to study variability of AGN
using a nearby sample, this has to be taken into account. In our case, a simple
removal of the nearby quasars is enough, since {\it none} of the other SF bins
are affected (again thanks to our bin independence).

It should also be noted that this turnover is not present (or at least has not
been detected) in the \gsdss-band. This is consistent with the notion that the
AGN is more dominant at shorter wavelengths, whereas a galaxy is usually much
redder.

In further discussions about the \rsdss-band SF curve, we have removed these
quasars (10\% of the total) from the sample. The \gsdss-band data-set is
unaffected.

\subsubsection{Light-curve asymmetry}
\label{LCasymmetry}

Asymmetries between the rising and falling parts of the light-curve can be
investigated by separating the variations in positive and negative variations
only (cf. \S~\ref{malmquistSF}). \citet{kawaguchi98} model various scenarios of
variability, each with different SF signatures. Their starburst (SB) model has a
very short rise time, followed by a long exponential decay (cf.  their Fig.~2).
They also consider an accretion disk instability (DI) model for which the
variations rise slowly, but fall off rapidly (basically in a saw-tooth like
pattern, cf. their Fig.~5). The SF curves derived from these light curves both
have significant asymmetries between the positive and negative variations.  They
are, predictably, of opposite nature: the fast-rise, slow-decline of the SB
model results in more SF signal in the positive variations, whereas the
slow-rise, fast-decline DI model has more signal in the negative variations.
This behavior can be intuitively understood in terms of the SF having typically
more variability signal at the fast changing part of the light curve compared to
the slowly changing part, for a given time-lag. In a sense the SF mirrors the
derivative of the light curve.

\begin{figure}[t]
\epsscale{1.0} 
\plotone{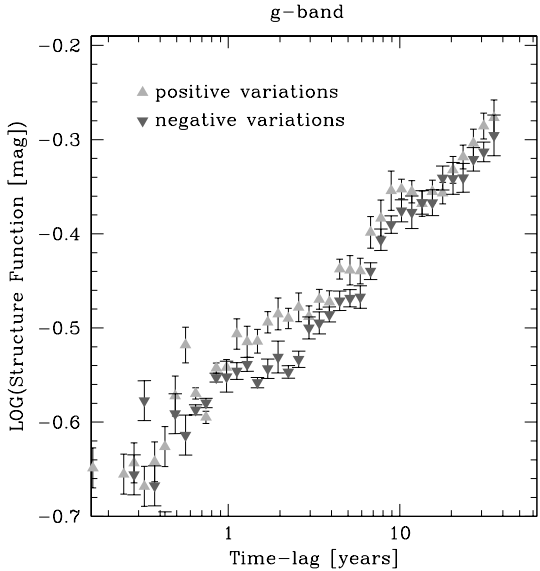}
\caption{The $g$-band Structure Functions for the quasars, separated into
positive-only (light-gray symbols) and negative-only (dark-gray symbols)
variations (see Fig.~\ref{SFposneg} for their definitions). Since the positive
SF curve has more signal than the negative one (which is the exact opposite of
the stellar case), it is a clear indication of asymmetric variability (i.e., the
rise and decline parts of the light-curve are not identical). The mean offset
along the $y$-axis is $\overline{O}=0.027\pm0.003$ (for time-lags beyond 1
year), which is equivalent to a 0.020 magnitude offset in rms.}
\label{SFQposneg}
\end{figure}

Figure~\ref{SFQposneg} shows this asymmetry for our data. We only show data for
the \gsdss-band because the SF offset between the \gsdss- and \rsdss-bands is of
the same order of magnitude as the asymmetry signal (cf. Fig.~\ref{SFquasars}),
so combining both bands is not helpful. The positive variations (light-gray
symbols) have more SF signal than the corresponding negative ones (they are
binned the same way in time-lag). This is a strong indication that the typical
quasar variations are not symmetric, and behave in a fast-rise, slow-decline
way. This is consistent with the modeled light curves we used in Paper~I (see
Fig.~5) and the SB-like light curves, but appears at odds with the inferred
behavior of the DI models. Furthermore, the asymmetry effect we see in the
calibration stars (cf. \S~\ref{malmquistSF}), and which we interpret as a sign
of Malmquist bias, works in the {\it opposite} sense. This increases the
significance of the disparity between the positive and negative variations for
our quasar sample. Indeed, based on modeled SF curves for which we force the
light-curves to be time-symmetric, we consistently measure a mean offset between
the positive and negative variations of $\overline{O}=0.000\pm0.003$ (based on
Eqns.~\ref{oline1} and \ref{oline2}).  For our actual quasar sample these values
are $\overline{O}=0.027\pm0.003$, which makes it significant at the 6.4$\sigma$
level.


Another trend that does not appear in our data is for the two SF curves to merge
beyond the typical time-scale of the variation. Both the SB and DI models of
\citet{kawaguchi98} display this behavior, whereas our SF curves remain offset
and parallel as function of time-lag. This is an indication that there does not
appear to be a preferred time-scale in our quasar sample. In \S~\ref{modelasym},
we will actually argue for a continuum of variation time-scales.

\subsubsection{Asymmetry implications for the micro-lensing scenario}

If variations are asymmetric in time, and either spend the least time rising or
declining in brightness, offsets will appear in the SF between the positive and
negative variations (e.g., Kawaguchi et al. 1998, and \S~\ref{modelasym}), given
enough of an asymmetry. This asymmetry signal persists under arbitrary
time-dilation corrections: a $(1+z)$ correction merely compresses the
time-scales, it does not alter the intrinsic shape of the light-curve. Neither
does it change the slope of the SF \citep[e.g.,][]{kawaguchi98,hawkins02}.  The
same statements are true for symmetric variations: there is no particular
$(1+z)$ correction that will induce an SF asymmetry signal if the variations are
intrinsically symmetric.

This implies that the asymmetry we detect in our SF is not an artifact of the
applied $(1+z)$ correction. In order to construct the SF in
Fig.~\ref{SFQposneg}, we used the quasar redshift itself to correct for
time-dilation, with the side benefit of improving our time-sampling. However, in
the case of micro-lensing, we would have to use the redshift of the lens, which
is unknown. This effectively prevents us from ever creating an SF that is
properly time-dilation corrected in the frames of the lenses. But {\it if} we
are solely interested in whether there is an asymmetry signal in the variations
or not, {\it we do not have to}. Our asymmetry signal persists, regardless of
the location of the lenses and its associated proper $(1+z)$ correction.

This asymmetry signal is at odds with the micro-lensing scenario, since these
events have to be symmetric in time (see for instance \citet{yonehara99} who
specifically modeled lensing of the AGN accretion disk). This strongly suggests
that micro-lensing cannot be the dominant cause of long-term variability in
quasars.  It quite likely contributes at some level, but not enough to define
the sample average behavior.

This contention is partly supported by \citet{zackrisson03}, who found that
micro-lensing cannot account completely for the observed long term variability,
based on its inability to explain the high number of large amplitude events and
the mean variability amplitude at low redshifts (where lensing is less likely).
Our results do constrain the level of micro-lensing a bit more though.

\subsubsection{Redshift and Absolute magnitude effects}


\begin{figure}[t]
\epsscale{1.0} 
\plotone{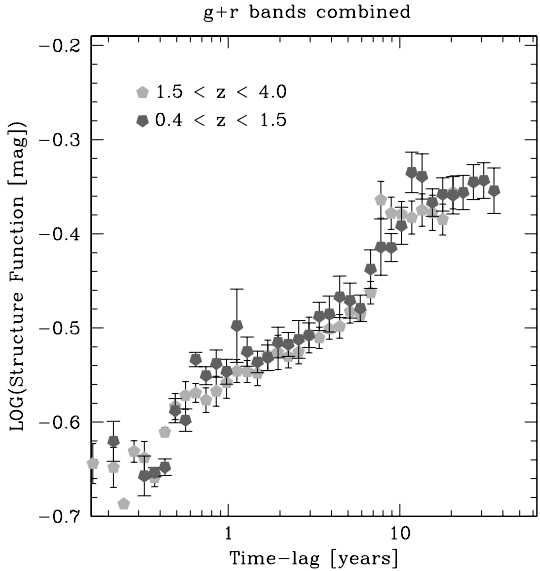}
\caption{Structure functions for low ($0.4 < z < 1.5$, dark-gray symbols) and high 
($1.5 < z < 4.0$, light-gray symbols) redshift quasars. The mean redshift for
each bin is 1.002 and 2.065, respectively. One would expect that the higher
redshift bin is more variable due to the intrinsically bluer part of the
spectrum that is probed (cf. Fig.~\ref{timelagVSz}). This is not the case,
however. The mean offset along the $y$-axis for time-lags beyond 1 year is
$\overline{O}=0.011\pm0.005$.}
\label{SFhighlowz}
\end{figure}

In the rest of the paper we assume that all the variations are intrinsic to the
quasar, and hence the $(1+z)$ time-dilation correction has to be applied.  If
one wants to consider both the time-dilation corrected and uncorrected cases,
one introduces another level of degeneracy between redshift, absolute
luminosity, and restframe spectral variability \citep{hawkins01}, which
needlessly complicates matters. Based on the results in the last section, we
feel confident that the variations are indeed source related.

This brings us to the first of the possible degeneracies: redshift and restframe
spectral variability. It is clear that sources vary more in the blue than in the
red \citep[cf. Fig.~\ref{SFquasars}, or][]{giveon99,trevese02,hawkins03}.  Any
trend in which higher redshift quasars become intrinsically less variable can be
offset against the increase of variability as the observed spectral range shifts
toward the blue with increasing redshift. Disentangling the redshift and
spectral variability contributions will be difficult, provided the higher
redshift SF curve lies above the lower redshift one. However, as is shown in
Fig.~\ref{SFhighlowz}, it is clear that this is not the case: the high-$z$
(light-gray symbols) SF curve lies {\it below} the low-$z$ one (dark-gray
symbols). This provides us with a solid lower limit on the trend that high-$z$
quasars are less variable than their low-$z$ counterparts.

We measure a mean offset along the $y$-axis of $\overline{O}=0.011\pm0.005$, for a
2.2$\sigma$ significance. Again, given the intrinsically bluer part of the
spectrum that is probed for the high redshift bin ($\overline{z}=2.065$, 18\,426
quasars) compared to the low redshift bin ($\overline{z}=1.002$, 18\,335
quasars), one expects the former bin to be more variable. That this is not the
case only increases the significance of the assessment that low redshift quasars
are more variable than high redshift ones.

The observed magnitude range for the quasars between $0.4 < z < 4.0$ is roughly
$18 < r < 21$ (cf. Fig.~\ref{brightHist}), and rather uncorrelated with
redshift. This in turn implies that the absolute \rsdss-band luminosity (after
taking a k-correction into account) of the sample increases as a function of
redshift. The two redshift bins of Fig.~\ref{SFhighlowz} are therefore also
separated in absolute luminosity. Since absolute luminosity is an intrinsic
property of the quasar (whereas redshift is not), it makes more sense to
investigate the variability as function of absolute luminosity. If one assumes
that the quasar variations are similar in an {\it absolute} sense, then the {\it
relative} variations are smaller for the intrinsically brighter objects.


\begin{figure}[t]
\epsscale{1.0} 
\plotone{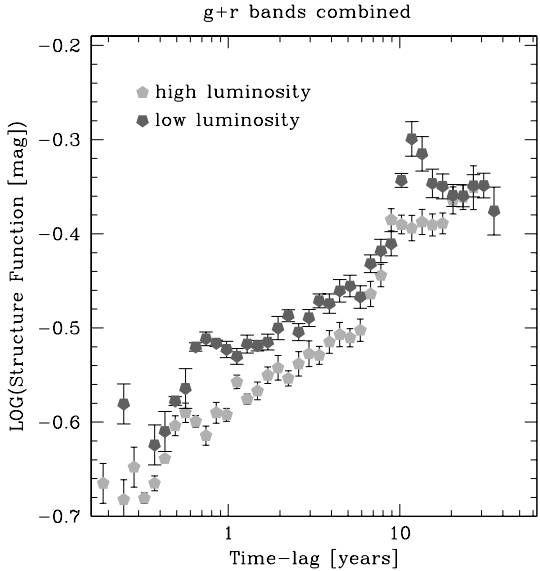}
\caption{Structure functions for low ($-20 > M_r > -24.32$, dark-gray symbols) and high 
($-24.32 > M_r > -29$, light-gray symbols) luminosity quasars. The subsample of
quasars is identical to the one for Fig.~\ref{SFhighlowz}, and $M_r = -24.32$
represents the mean absolute \rsdss-band luminosity. The offset between the low
and high luminosity SF curves is best explained along the $y$-axis, and not the
$x$-axis. In other words, the relative variability in low luminosity quasars is
larger than in high luminosity quasars, while it does not have a different
preferred time-scale.}
\label{SFhighlowl}
\end{figure}

Figure~\ref{SFhighlowl} shows this trend. The subsample of the quasars with
redshifts $0.4 < z < 4.0$ has been divided up into two bins: all the quasars
fainter, and brighter than $M_r = -24.32$. This value is the mean absolute
luminosity of the subsample\footnote{We adopt H$_{\rm o}$=71 km s$^{-1}$
Mpc$^{-1}$, $\Omega_{\rm M} = 0.3$, and $\Omega_\Lambda = 0.7$ throughout this
paper.} Note that for the low-luminosity quasars (the dark-gray symbols) the
photometry quality degrades toward higher redshifts, which expresses itself as
an artificial increase in the SF signal (especially for the time-lags between 10
to 20 years).  The GSC2 photometry is less affected by this, resulting in a
nicer signal. The SF curves are offset significantly (and more so than in
Fig.~\ref{SFhighlowz}) between time-lags of about a year to $\sim8$ years.

Based on Figs.~\ref{SFhighlowz} and \ref{SFhighlowl}, we conclude that
high-luminosity quasars vary less than low luminosity quasars. This supports
earlier similar findings by, e.g., \citet{hook94,trevese94,cristiani96}, but
contradicts for instance, \citet{giallongo91}. These authors argued that the
absence of the trend for higher luminosity quasars to be less variable points
toward a scenario in which the object varies as a single (coherent) source,
rather than a flaring sub-unit. Given our results, we have to conclude the
opposite, namely that quasars tend to vary incoherently, and do so with a
limited magnitude range of the flaring sub-units.

\citet{garcia99} tried to explain this within a supernova context
\citep[e.g.,][]{aretxaga97} in which the time separation between outburst (with
total energies of up to $10^{50}$ ergs) are distributed in a Poissonian way.
However, the resulting SF curves based on supernovae are too steep
\citep[cf.][]{kawaguchi98,hawkins02}, and are effectively ruled out. In Paper~I,
we applied a similar shot-noise outburst distribution to model the SF curve, but
used a broader (more generalized) exponential light-curve. In \S~\ref{Modeling},
we will get back to this issue.

\section{Modeling the Structure Function}
\label{Modeling}


\begin{figure}[t]
\epsscale{1.0} 
\plotone{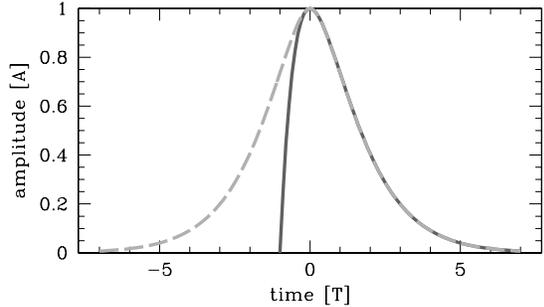}
\caption{Model light-curves, plotted against time. The asymmetric light-curve is 
plotted in dark-gray, and is given by Eqn.~\ref{LCeqnASYM}. Its symmetric
counterpart is plotted in light-gray (Eqn.~\ref{LCeqnSYM}).  The units of time
and amplitude have been normalized by $T$ and $A$, respectively. Note that the
signal of the asymmetric curve at times $t < -T$ has been set to zero. }
\label{modelCurves}
\end{figure}

In this section we will expand on the modeling done in Paper~I. The limited
sample size and the resulting noisy SF curve did not allow for accurate fitting
in that paper. With the present data, however, we are in a position to
investigate this further\footnote{Our modeling setup does not assume any
a-priori information about how the SF was constructed. For all practical
purposes, we could have been fitting to the Hawkins (2002) data-set, or a
straight line fit to those data.}. The main modeling result from Paper~I was
that we could approximate the observed SF curve by assuming the following: 1)
the typical quasar undergoes periodic outbursts with a decaying time-scale of
$\sim 2$ years, and 2) these outbursts occur on a typical time-scale of $\sim
200$ years. As we will see in the next few sections, this simple picture is not
correct.

First, let us begin describing the modeling setup we used. Like in Paper~I, we
assume a typical outburst can be described by a canonical exponential function
(Eqn.~\ref{LCeqnASYM}).  Curves of these types can be used to model various
possible outburst scenarios, ranging from short time-scale supernovae to longer
time-scale accretion disk instabilities \citep[see, e.g.,][and references
therein]{kawaguchi98}. In addition to Paper~I, we also introduce a symmetric
version of Eqn.~\ref{LCeqnASYM}, given in Eqn.~\ref{LCeqnSYM}. We need both
curves to adequately investigate the observed SF asymmetry (cf.
\S~\ref{modelResults}).

For each equation, $A$ represents the amplitude of the outburst (in magnitudes),
$e$ is the normalization constant so that the peak has an amplitude $A$, and $T$
is the exponential half-life time (in years). The time parameter $t$ has been
chosen such that the peak of the outburst occurs at time $t=0$. The normalized
light-curves have been plotted in Fig.~\ref{modelCurves} to make this a little
clearer. The asymmetric curve (plotted in dark gray) is the exact same as in
Paper~I, and the symmetric case is identical to the asymmetric one for times $t
> 0$.

\begin{equation}
L_{\mbox{\small asym}}(t) = e A \left(\frac{t+T}{T}\right)  e^{-(t+T)/T}
\label{LCeqnASYM}
\end{equation}

\begin{equation}
L_{\mbox{\small sym}}(t) = e A \left(\frac{|t|+T}{T}\right)  e^{-(|t|+T)/T}
\label{LCeqnSYM}
\end{equation}

\noindent Note that Eqn.~\ref{LCeqnASYM} is only valid for times $t > -T$. The
quantities $A$ and $T$ are free parameters in the model. For each of these
points in phase-space, a quasar composite light-curve is constructed covering a
large time period (typically 10\,000 years). This period is filled up with
outbursts based on $A$ and $T$, separated in time by a characteristic outburst
time-scale $P$. Like in Paper~I, the probability of an outburst {\it not
occurring} within time $t$ is given by

\begin{equation}
\mbox{Prob($t$)} dt = \left(\frac{1}{P}\right) e^{-t/P} dt
\label{shotnoise}
\end{equation}

\noindent which is also known as a shot noise model of variability \citep[see,
e.g.,][]{lochner91}. Once we have constructed such a canonical quasar
light-curve (based on the values of $A$, $T$, and $P$), we generate a database
of measurements which is identical to the actual one for our quasars.  So we use
the very same redshifts (all 41\,391), and with on average 4 measurements per
quasar (from this canonical light-curve), this results in about 250\,000
permutations. This is close to the actual values (170\,000 for the \gsdss-band,
and 130\,000 for the \rsdss-band). To each, randomly sampled, ``measurement'' we
add white noise that matches the actual measurement uncertainties. We adopt the
conservative values of $\sigma_{\mbox{\tiny SDSS}}=0.04$, $\sigma_{\mbox{\tiny
GSC}}=0.12$, and $\sigma_{\mbox{\tiny POSS}}=0.21$ magnitudes. The error in the
magnitude differences are then the rms-values of the appropriate $\sigma$'s. The
first two values are in part set by the SF noise plateau at very short
time-scales (which do not contain any POSS data), and the $\sigma_{\mbox{\tiny
POSS}}$ value is based on the stellar SF plateau in \S~\ref{stellartypedep},
which can be considered an upper limit (\citet{sesar04} quote a $\sigma=0.15$
mag). The artificial SF curve which is based on these data can then directly be
compared to the actual one.

The model SF curve is sampled on the exact same time-lag binning as the actual
one, allowing for a simple comparison. There is, however, a complication. The
parameter combination of $A$ and $P$ turns out to be degenerate. If one
increases the amplitude $A$, and at the same time makes outbursts rarer by
increasing the typical time-scale $P$, it results in the same SF curve (which is
still dependent on $T$) as for smaller values of $A$ and $P$.  The sole thing
that discriminates between the two scenarios is that the error-bars in the high
$A$, high $P$ case are much larger than in the low $A$, low $P$ case. This can
be understood in terms of what defines the typical quasar behavior: in the high
$A$, high $P$ case, most quasars will not vary, except for a small subset. This
introduces large SF variations, depending on whether such measurements are
present in the bin or not. In the low $A$, low $P$ case, a lot of quasars vary,
but do so at low amplitude. The relative sample variations are small, and most
time-lag bins are made up of more homogeneous measurements. Remember, the
error-bars in the SF curves are based on the rms within a set of time-lag bins
(we bin bins, see Paper~I for more details).

So, instead of a simple $\chi^2$ value, our goodness-of-fit is defined by the
sum of the $\chi^2$ value and the rms in the difference between the sizes of the
model and actual error-bars.  Their relative values are normalized such that for
a good fit each contribution has about equal weight.

The best fitting values of $A$, $T$, and $P$ (or sets thereof in cases where we
fit multiple components) are found by exploring the phase-space extensively.
Since this space typically does not have any steep gradients, we opted for a
slow-cooling simulated annealing code, specifically developed for this purpose.
It consistently converged to the same solutions, provided we cooled slow enough.

\subsection{Modeling Results}\label{modelResults}


\begin{figure}[t]
\epsscale{1.0} 
\plotone{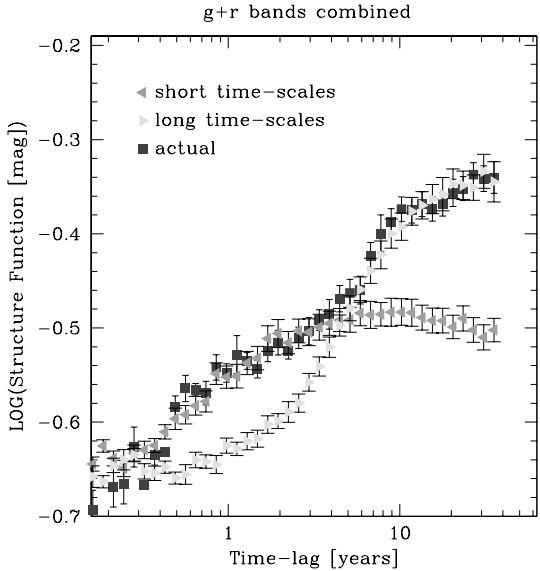}
\caption{Plot showing the inability to fit the actual SF curve with
one single set of $A$, $T$, and $P$ values. Unlike Paper~I, where we were able
to fit the much poorer data with a single set, this is not possible with the
current data.  A combination of short- and long-term variability is needed. In
this case, separate fits have been made to the actual SF curve: one for
time-lags below 5 years, and one for time-lags beyond 5 years.}
\label{SFqmodel}
\end{figure}

In the first subsection, we will restrict the modeling to the symmetric
light-curve functions (Eqn.~\ref{LCeqnSYM}). After that we are also considering
the asymmetric light-curves of Eqn.~\ref{LCeqnASYM}.

\subsubsection{Inability to fit single component}

In Paper~I we were able to fit the observed SF curve adequately by a single set
of $ATP$ values, suggesting a possible characteristic variability time-scale /
mechanism common to all quasars. The data quality, however, was such that this
could not be put on sure footing. With the current high quality data, this is
any easy thing to test. As can be seen in Fig.~\ref{SFqmodel}, we cannot fit the
observed SF curve with a single component.  The two model curves, one fitting
the short ($<5$ years) time-lags, and one fitting the long ($>5$ years)
time-lags illustrate the problem. In order to generate enough signal at the
short time-lags, the model light-curve has to have a small half-life $T$. The
actual values for the fit are: $A=0.4, T=0.29, P=4$, with $A$ in magnitudes,
and $T$, $P$ in years. However, it is also clear that the SF curve belonging to
this light-curve does not have any signal increase beyond a few $T$; it
effectively levels off after $\sim 3$ years and starts deviating from the actual
SF curve significantly. On the other hand, SF model curves that do fit the
longer time-lags (based in this case on light-curves with a half-life $T=2.24$
years, $A=0.5$, and $P=18$), do not have any significant SF signal on
time-scales much shorter than their $T$ value.

We therefore need at least two variability components. This refutes the notion
from Paper~I that there is a single preferred variability time-scale common to
all quasars. It also implies that there may be a continuum of variability
time-scales. Our best fitting two component fit (to the SF curve plotted in
Fig.~\ref{SFqmodel}) has values of: $A=0.9$, $T=0.27$, $P=23$, and $A=0.5$,
$T=3.67$, $P=42$. Again, due to the degeneracy between $A$ and $P$, their values
are not very well constrained. Longer periods $P$ are not excluded.  However, we
did try to make the error-bars on the model SF curve as much like the actual
data as possible.

Even fitting with more components did not alter the shortest time-scales. Most
of the shortest $T$ values grouped around 0.2 year. Our data have some
sensitivity to these short time-lags, but not an awful lot. About 2\%\ of the
available time-lag measurements are shorter than 0.2 years. This half-life of up
to a few months is comparable to high-redshift supernovae timescales \citep[see,
e.g.,][]{barris04}. The longer time-scales needed to fit the SF curve beyond a
few years are more in line with disk instability models of large accretion disks
around supermassive black holes \citep[e.g.,][]{kawaguchi98}. It is therefore
not immediately clear what, if any, mechanism dominates.

\subsubsection{Modeling SF asymmetries}
\label{modelasym}


\begin{figure}[t]
\epsscale{1.0} 
\plotone{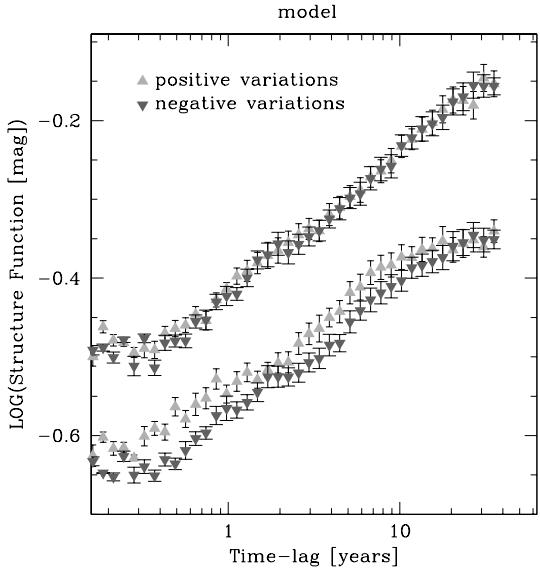}
\caption{The asymmetry behavior of a particular 3 component model (with
half-life values of $T=0.34, T=2.90, T=6.94$, bottom set of data-points), using
a light-curve described by Eqn.~\ref{LCeqnASYM}. Even though the combined SF
(with both positive- and negative-only variations) fits the observed SF very
well, the asymmetry is a bit different from the actual case
(Fig.~\ref{SFQposneg}): the magnitude of the asymmetry slightly smaller
($\overline{O}=0.022\pm0.003$), and the ``nulls'' at $\sim$2 and $\sim$30 year
time-lags are not observed either. By comparison, the top set of points
(vertically offset by 0.15 for clarity), represents the symmetric case
(Eqn.~\ref{LCeqnSYM}) for exactly the same set of parameters and initial random
seed. The asymmetry is measured to be $\overline{O}=0.003\pm0.003$.}
\label{SFMposneg}
\end{figure}

As we have seen in \S~\ref{LCasymmetry}, our quasar SF curve displays
significant asymmetry, with a mean offset of $\overline{O}=0.027\pm0.003$. This
is not compatible with our symmetric fits from the previous section, which
consistently resulted in offsets of $\overline{O}\approx0.000$, to within the
$0.003$ uncertainty (cf. Fig.~\ref{SFMposneg}, top curve). Hence, we need to
introduce some measure of light-curve asymmetry into our modeling.  For this, we
use Eqn.~\ref{LCeqnASYM} as the functional form of the light-curve.

The fitting method is the same as for the symmetric light-curve case, and the
results are reasonably comparable. Given the shape of the light-curves (cf.
Fig.~\ref{modelCurves}), one expects to find larger values of $T$ for the
asymmetric case than for the symmetric one. This is because the latter
light-curve has variations at longer time-scales than the asymmetric one for
identical values of $T$.

We find for the asymmetric 2 component case: $A=0.8$, $T=0.45$, $P=23$, and
$A=0.5$, $T=5.43$, $P=42$, which has indeed larger values of $T$ than the
symmetric case.  It also carries the asymmetry signal we are interested in. A
typical resulting SF is shown in Fig.~\ref{SFMposneg}, separated into negative-
and positive-only variations. The SF curves are clearly separated (mean offset
$\overline{O}=0.022\pm0.003$ for this case). However, there are a few things
that are not entirely consistent with the observed data. Aside from the slightly
smaller value of the asymmetry signal $\overline{O}$, it appears that for
time-lags somewhat shorter than $T$, the SF does not have an asymmetry signal.
This is not observed in the actual data (cf. Fig.~\ref{SFQposneg}).  Since these
nulls are associated with the few discrete time-scales $T$, adding in more
intermediate $T$'s will smooth out and remove the nulls. This suggests that the
intrinsic quasar variations have a more continuous distribution of
$T$'s. Indeed, our models with more than four components tend not to have these
nulls.

The data, and our modeling effort, do not allow for a precise characterization
of the magnitude of the light-curve asymmetry. It also does not allow for the
isolation of particular variability time-scales, and is more compatible with a
scenario in which quasars can vary with a continuous distribution of
time-scales. One thing that is clear, though, is that the variations are
asymmetric, and are of the fast-rise, slow-decline type.

\subsubsection{Estimating intrinsic SF slope}\label{intslope}


\begin{figure}[t]
\epsscale{1.0} 
\plotone{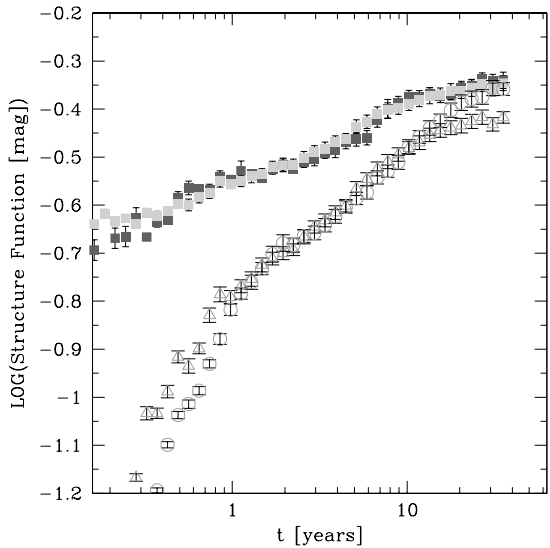}
\caption{SF slope changes due to removal of measurement noise. The top two sets
of points represent the actual SF and best fitting 3 component model (same as in
Fig.~\ref{SFMposneg}). Note that the y-axis scale has been expanded. The slope
of these curves is $\alpha=0.15\pm0.01$. The bottom 2 sets of points represent
the SF for the exact same 3 component model, but with the magnitude measurement
noise for the SDSS, GSC, and POSS surveys set to zero. The triangles and circles
are for the asymmetric and symmetric light-curves respectively. The slopes for
both SFs are $\alpha=0.30\pm0.01$, at least for the flat part between time-lags
of 1 to 20 years.}
\label{SFnonoise}
\end{figure}

The theoretical slope calculations in \citet{kawaguchi98} do not include a white
noise component, as their data quality for the individual sources was good
enough. We, however, do have to include a white noise term in our modeling to
make it agree with the actual data better. This does have the side-effect of
lowering the SF slope (as noted in Paper~I), which potentially might render the
slope comparison suspect. To this end, we have generated ``noise-less'' SF
curves derived from a well fitting ``noise-included'' SF curve. The results have
been plotted in Fig.~\ref{SFnonoise}. The top set of data-points are the actual
SF with the best fitting (3 component in this case) SF model. The slope for both
curves is measured to be $0.15\pm0.01$, which, as discussed in \S~\ref{sfslope},
is much shallower than any of the \citet{kawaguchi98} values. If we remove the
measurement noise components, we end up with the two bottom sets of SF curves (a
symmetric and an asymmetric one). Note that even though there is no measurement
noise present, there will always be a range of magnitude differences for a fixed
time-lag (unless there is no variation), which explains the presence of
error-bars.

The slope for these curves is measured to be $\alpha=0.30\pm0.01$, which indeed
is steeper. The fall-off below time-lags of a year is due to the fact that the
actual SF (top curve in the figure) does not have any sensitivity to these short
time-scales, as they are effectively masked by the measurement noise. Since we
just de-noised the fit to the actual SF, this lack of short time-scale signal
becomes apparent. The quoted slope, therefore, is only for the SF curve beyond
time-lags of one year.

While the slope of the SF has indeed steepened a bit in the noise-less case, it
is not enough to change the assessment of \S~\ref{sfslope} in a significant
way. It is still too shallow for either the SB or DI models, but is now
marginally steeper than the micro-lensing slope.

\section{Discussion and Summary}
\label{Summary}

Our results on the quasar SF strongly suggest, for the first time, that most of
the long-term variations are intrinsic to the quasar itself. Micro-lensing by
objects along the line-of-sight to the quasar, or even in the quasar host galaxy
itself, is not a viable explanation of the long-term variability in general. 

This result is mainly based on the observed asymmetry in the SF, indicative of a
fast-rise, slow-decline type of variability (cf. \S~\ref{LCasymmetry}). The
significance of the observed asymmetry is enhanced by the Malmquist signal,
which works in the opposite sense. If our measured asymmetry would have had the
same sign as the Malmquist bias, one would be hard pressed to ascribe any of
that asymmetry to the intrinsic quasar variability behavior, and the data would
have been consistent with the symmetric variations needed in the micro-lensing
scenario. This is, however, not the case. The formal statistical significance of
the asymmetry is $6\sigma$.

We have put some other results on a more secure footing as well. First, no
obvious turnover has been detected in the quasar SF, which indicates that there
is no upper preferred variability time-scale (smaller than a few decades). Our
results are consistent with a continuum range of variation time-scales. This is
based on the absence of a turnover, as well as on the near constant offset
between positive- and negative-only variations in Fig.~\ref{timelagVSz}. If we
model asymmetry based on a few preferred time-scales $T$, we find almost no
asymmetry signal at timescales of a few times $T$ (cf. Fig.~\ref{SFMposneg}),
something clearly absent in the real data. A more or less constant offset in the
model can be achieved with a larger number of components (4 or more), all with
time-scales $T$ less than 10 years. This opens up the door to any number of
components, since as far as we know there is no observed turnover beyond
$\sim$40 years which would set an upper bound.

Second, the magnitude of the quasars variability is a clear function of
wavelength: variability increases toward the blue part of the spectrum. This
confirms previous observations by, e.g., \citet{giallongo91,giveon99,trevese02,
hawkins03}, and is consistent with both the starburst (SB) and the disk
instability (DI) model of variability. However, based on the model SF slopes by
\citet{kawaguchi98}, the SF slopes for the SB model $(0.74<\alpha<0.90$) are
even more inconsistent with our ``noise-less'' modeled value of
$(\alpha=0.30\pm0.01)$ than the DI values of $(0.41<\alpha<0.49)$. Clearly, some
modifications need to be made before the models can adequately describe the
observations.

Third, high-luminosity quasars vary less than low-luminosity quasars. This is
consistent with a scenario in which variations have a limited absolute
magnitude, and variations are due to sub-components instead of coherent
variation of the AGN \citep[e.g.,][]{garcia99}. These sub-components can be
interpreted as either individual supernovae, or discrete flares due to disk
instabilities.

In summary, all the data presented here lead to the conclusion that quasar
variability is intrinsic to the source, is caused by chromatic outbursts /
flares with limited luminosity range and varying time-scales, and with an
overall asymmetric light-curve. Currently, the model that best explains this
observed behavior is based on accretion disk instabilities. However, given the
existing discrepancies between the SF slopes of the model and observations, some
reservations are still in place.

\acknowledgments

We like to thank the referee for comments that helped improve the paper.  WDVs
work was performed under the auspices of the U.S. Department of Energy, National
Nuclear Security Administration by the University of California, Lawrence
Livermore National Laboratory under contract No.  W-7405-Eng-48.  The authors
also acknowledge support from the National Radio Astronomy Observatory, the
National Science Foundations (grants AST 00-98259 and AST 00-98355), the Space
Telescope Science Institute, and Microsoft.

\end{document}